\documentclass[10pt,conference]{IEEEtran}
\usepackage{mathptmx} 

\usepackage{fancyhdr}
\usepackage[normalem]{ulem}
\usepackage[hyphens]{url}
\usepackage[sort,nocompress]{cite}
\usepackage[final]{microtype}
\usepackage[keeplastbox]{flushend}
\usepackage[bookmarks=true,breaklinks=true,letterpaper=true,colorlinks,citecolor=blue,linkcolor=blue,urlcolor=blue]{hyperref}

\usepackage{multirow}
\usepackage{stfloats}

\usepackage{amssymb}
\usepackage{threeparttable}
\usepackage{tikz}
\usepackage{enumitem}
\usepackage{array}
\usepackage{multirow}
\usepackage{authblk}

\usepackage{subfigure}

\newcommand{\micro}[1]{\textcolor{black}{#1}}

\newcommand*\circled[1]{\tikz[baseline=(char.base)]{%
            \node[shape=circle,draw,inner sep=1pt] (char) {#1};}}

\newcommand{\rebuttal}[1]{\textcolor{black}{#1}}
\newcommand{\hpca}[1]{\textcolor{black}{#1}}

\title{Privacy Preserving In-memory Computing Engine}

\author[1]{Haoran Geng}
\author[2]{Jianqiao Mo}
\author[3]{Dayane Reis}
\author[1]{Jonathan Takeshita}
\author[1]{\\Taeho Jung}
\author[2]{Brandon Reagen}
\author[1]{Michael Niemier}
\author[1]{Xiaobo Sharon Hu}

\affil[1]{University of Notre Dame}
\affil[2]{New York University}
\affil[3]{University of South Florida}
{
    \makeatletter
    \renewcommand\AB@affilsepx{: \protect\Affilfont}
    \makeatother

    \affil[ ]{Email}

    \makeatletter
    \renewcommand\AB@affilsepx{, \protect\Affilfont}
    \makeatother

    \affil[1]{\{hgeng, jtakeshi, tjung,mniemier, shu\}@nd. edu}
    \affil[2]{\{jm8782, bjr5\}@nyu.edu}
    \affil[3]{\{dayane3\}@usf.edu}
}

\begin{document}
\maketitle
\pagestyle{plain}

\begin{abstract}
Privacy has rapidly become a major concern/design consideration. Homomorphic Encryption (HE) and Garbled Circuits (GC) are privacy-preserving techniques that support computations on encrypted data. HE and GC can complement each other, as HE is more efficient for linear operations, while GC is more effective for non-linear operations. Together, they enable complex computing tasks, such as machine learning, to be performed exactly on ciphertexts. However, HE and GC introduce two major bottlenecks: an elevated computational overhead and high data transfer costs. This paper presents PPIMCE, an in-memory computing (IMC) fabric designed to mitigate both computational overhead and data transfer issues. Through the use of multiple IMC cores for high parallelism, and by leveraging in-SRAM IMC for data management, PPIMCE offers a compact, energy-efficient solution for accelerating HE and GC. PPIMCE achieves a 107$\times$ speedup against a CPU implementation of GC. Additionally, PPIMCE achieves a 1,500$\times$ and 800$\times$ speedup compared to CPU and GPU implementations of CKKS-based HE multiplications. For privacy-preserving machine learning inference, PPIMCE attains a 1,000$\times$ speedup compared to CPU and a 12$\times$ speedup against CraterLake, the state-of-art privacy preserving computation accelerator. 

\end{abstract}

\section{Introduction}
\label{sec:introduction}
Privacy-preserving computation (PPC), where computations are performed directly on encrypted data, is a 
  solution for providing security and privacy in modern systems. However, PPC techniques typically incur extremely high computation costs. For example, machine learning (ML) inference with encrypted data can be 10,000$\times$ to 100,000$\times$ slower than plaintext \cite{HE_F1,craterlake,HE_cheetah,takeshita2022heprofiler}. Thus, there is a great need for solutions that mitigate the performance overhead of PPC.

One of the most promising PPC techniques is homomorphic encryption (HE) \cite{gentry09_FHE, BGV, B/FV}. HE allows computations to be performed directly on ciphertexts. In cloud computing, HE protects clients' privacy, as data remains encrypted during server side computation. While HE strengthens security and privacy, it introduces substantial computation overhead due to (1) high volume of data generated by large ciphertexts (i.e., \textit{ciphertext expansion}) \cite{verbauwhede_he_data, CIM_HE}, and (2)
expensive bootstrapping operations, especially for deep neural networks (DNN) that require many nested multiplications (e.g., \cite{bourse_2018_DNN_HE}). Additionally, many HE schemes lack support for non-linear operations, including Brakerski/Fan-Vercauteren (B/FV) \cite{B/FV}, Brakerski-Gentry-Vaikunathan (BGV) \cite{BGV} and Cheon-Kim-Kim-Song (CKKS) \cite{CKKS}, which can impact DNN accuracy \cite{HE_accuracy_dropping}.

Garbled Circuits (GC) are an alternative PPC technique that can efficiently support non-linear functions. GC can logically operate on encrypted binary data, allowing arbitrary computations. Numerous advancements have contributed to optimizing the performance of GC-based applications \cite{Freexor,GC_TINY,HalfGate}.
State-of-the-art (SOTA) private machine learning protocols~\cite{Delphi, GAZELLE,Relu_dropping,GC_data_transfer} use HE for linear operations and GC for non-linear to achieve high accuracy.
However, previous work also shows that GC can suffer from high computational costs  \cite{GC_ReLu} and large client-server communication overheads \cite{HE_F1,craterlake}.

Hardware accelerators exist for both HE \cite{HE_F1,BTS,craterlake,HE_cheetah} and GC~\cite{GC_overlay,Maxelerator,FASE,HAAC}. Although these accelerators yield high performance, they also suffer from large data transfer overheads \cite{GC_data_transfer,HE_F1}. Recent research suggests that in-memory computing (IMC) is a viable solution \cite{CryptoPIM,CIM_HE,CIM_HE_SAC,Xpoly}. IMC has been proposed as an architectural solution to overcome latency and energy overheads both associate with data transfer \cite{sebastian17,why_imc1,why_imc2}. With an IMC architecture, a subset of logic, arithmetic, and memory operations associated with given tasks are performed in memory (without transfers to/from a processor). IMC exploits the large internal memory bandwidth to achieve parallelism, which reduces latency and saves energy due to fewer external memory references.

Existing PPC accelerators have only targeted HE or GC, making stand-alone solutions suboptimal for certain PPC tasks. 
For example, in privacy-preserving machine learning (PPML) inference, HE cannot easily support non-linear operations like ReLUs while maintaining high accuracy~\cite{GC_data_transfer,cryptonite}.
Therefore, existing HE accelerators must replace the non-linear activation functions in ML algorithms with HE-friendly operations using methods such as polynomial approximation.  These HE-friendly activation functions cause a significant drop in accuracy \cite{HE_accuracy_dropping}. Alternatively, while a GC accelerator can accelerate ReLU functions, it can be extremely inefficient for matrix-vector multiplication in ML algorithms.


Using a combination of HE and GC (e.g., \cite{GAZELLE}) allows the execution of PPML tasks {\it without any loss of accuracy}. Our experiments also show that the combined HE+GC protocol offers less latency for PPML tasks than a pure HE approach due to HE bootstrapping overheads. Therefore we introduce the Privacy Preserving In-memory Computing Engine (PPIMCE), an IMC architecture designed to accelerate HE and GC in a single, unified hardware platform. In PPIMCE, we leverage the high parallelism, high throughput, low data transfer time, and low energy usage offered by IMC to overcome the performance and data transfer bottlenecks in HE and GC.

\hpca{ The key insight behind our approach is the use of an in-SRAM IMC accelerator for executing HE and GC, substantially mitigating data transfer costs between on-chip memory and processing units. The PPIMCE system utilizes specialized IMC cores designed to perform a range of operations tailored to the combined use of HE and GC. These cores, strategically placed near memory arrays, optimize data transfer and surpass traditional ASICs in efficiency. One significant challenge we confront is integrating HE and GC, two fundamentally distinct algorithms, into a single system. To tackle this, we leverage our IMC cores' proficiency in handling basic logical and arithmetic operations. Additionally, we employ a scheduler that efficiently coordinates these operations, facilitating the concurrent execution of HE and GC tasks within the system.  Our key contributions can be summarized as follows:}




\begin{itemize}[noitemsep]

    \item PPIMCE is the first hardware accelerator based on IMC that can execute all essential operations to support HE and GC with high performance.
    \item The mapping and scheduling scheme in PPIMCE enables the high-performance realization of HE and GC.
    \item A thorough evaluation shows PPIMCE outperforms CPU, GPU, and SOTA PPC accelerators in latency, area, and power across diverse benchmarks.

\end{itemize}

Our experimental results, detailed in Section \ref{sec:Experimental and Evaluation}, show PPIMCE's superior performance. We observe a 100$\times$ latency improvement over CPU-based GC and a remarkable 1,500$\times$ and 800$\times$ speedup over CPU and GPU in CKKS-based HE multiplications. Furthermore, PPIMCE surpasses existing PPML solutions, offering a 1,000$\times$ speedup over Gazelle and up to 130$\times$ over the latest PPC accelerators, all within a compact 138.8$mm^{2}$ area and just 9.4$W$ average power consumption.

\section{Motivation and Challenges}
\label{sec:Motivation}

\hpca{This section highlights our primary motivations: using GC for non-linear layers in PPML inference to avoid expensive bootstrapping and exploiting IMC's performance against ASIC for data-intensive applications like HE and GC. The main challenge in designing PPIMCE is the integration of two fundamentally distinct algorithms, HE and GC, into a singular hardware architecture.}

\subsection{HE+GC vs. HE-only protocol for PPML inference}

\label{sec:HE+GC_VS_HEonly}

\hpca{There are two primary methods for PPML inference: (1) exclusively using HE\cite{lola,HE_resnet_20,HEProfiler}, or (2) using mixed protocols that use HE for linear and GC for non-linear functions\cite{GAZELLE,Delphi}. The HE-only approach outsources all computations to the server, incurs low communication costs, but significantly increases latency due to HE bootstrapping. In contrast, while the mixed HE+GC protocols require increased communication, they can avoid bootstrapping and reduce computation latency. We aimed to compare the communication latency associated with GC with the bootstrapping demands of an HE-only strategy.}

 \hpca{The HE-only runtime includes latencies from HE linear computations and bootstrapping. For non-linear functions, we assume degree-6 polynomial approximation for ReLU activations~\cite{HE_non_linear}.
 The HE-only computation is implemented and profiled with the SOTA CKKS library HEAAN~\cite{HE_CPU_HEAAN}.}

\hpca{In the HE+GC method, we follow the Gazelle framework\cite{GAZELLE}. Runtime consists of HE computations for linear functions and GC Garbler computations for non-linear functions. To support HE tasks, we use the SEAL library\cite{SEAL} (SEAL and HEAAN show comparable performance\cite{HEProfiler}). The emp-tool library\cite{emp-tool} is employed for GC computations. To account for communication latencies, we considered the communication latency of GC under various communication protocols, assuming bandwidths of 2G\cite{2G} to 5G\cite{5G} network. This allows us to comprehensively assess the GC delay on the server side.}

\begin{figure}[t]
  \centering
  \resizebox{.98\columnwidth}{!}{\includegraphics{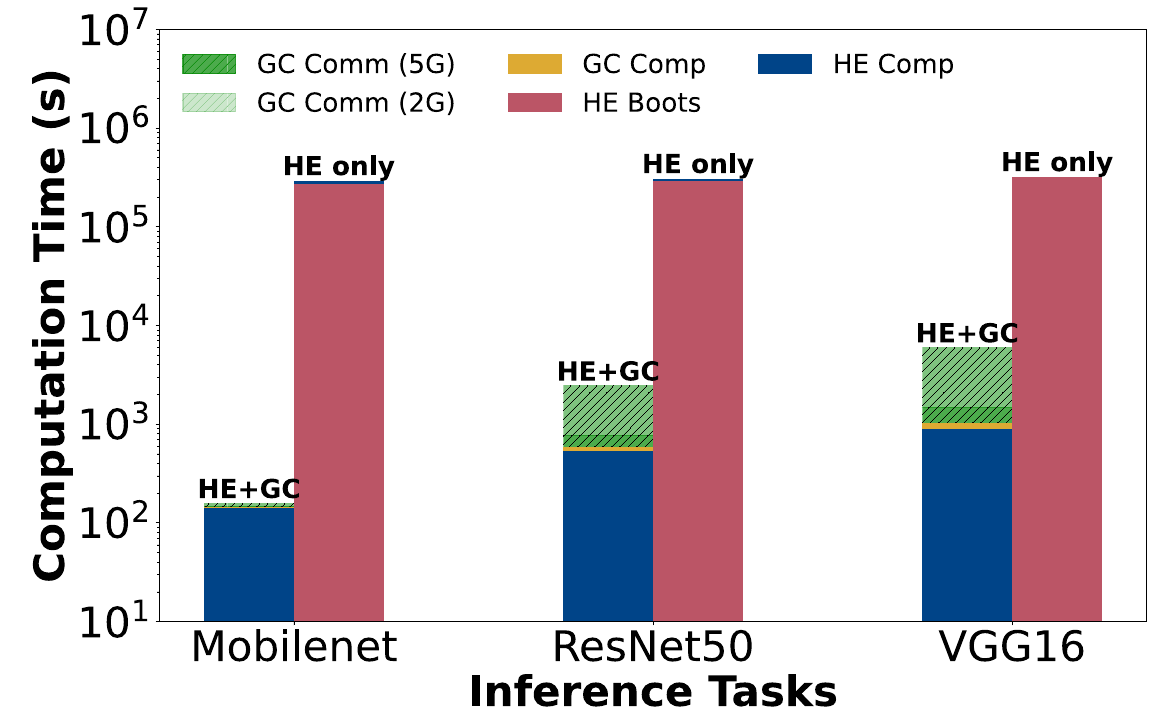}}
  \caption{Analysis of CPU computation time in HE-only and HE+GC approaches for PPML inference.}
  \label{fig:heaan_vs_gazelle}
\vspace*{-5mm}
\end{figure}

\hpca{Figure \ref{fig:heaan_vs_gazelle} illustrates the time difference between the HE-only and HE+GC approaches. In the HE-only approach, bootstrapping consumed over 95$\%$ of the computation time, resulting in inference times of several days. Conversely, the HE+GC approach avoids bootstrapping, thus reducing the total inference time to mere hours, as GC computation is considerably faster. Despite the communication cost associated with GC, reduced bootstrapping overhead from the HE+GC approach can improve efficiency. These savings underscore PPIMCE's advantage, allowing it to outperform accelerators using only HE on PPML inference due to its effective execution of both HE and GC operations (See Section \ref{sec:PPML_inference}).}

\subsection{Advantages of IMC}
\label{sec:Advantages_of_IMC}

\hpca{As emphasized in \cite{computecache}, one major bottleneck in data-intensive applications like HE and GC is the substantial data movement overhead. This issue arises from the need to move large data between memory and computing units. For instance, to ensure a security level of 256 in HE, a single ciphertext size is 16MB, as the ciphertext is represented as a high-degree polynomial. A single convolutional layer in PPML inference might require as much as 256MB of ciphertext\cite{HE_F1}. Similar issues are encountered with a GC approach as each bit of plaintext is encrypted into 128-bit secret labels. Thus, compared to non-GC solutions, GC involves the transfer of more than 128 times the volume of data from memory to a computing unit.}

\hpca{IMC can alleviate this data movement overhead. IMC architectures can efficiently perform bitwise and arithmetic operations inside the memory, significantly improving efficiency for data-intensive applications like HE and GC. The potential of IMC to revolutionize hardware accelerators for such applications has garnered interest from academic circles \cite{verma19, sebastian2020memory}, government agencies \cite{darpacall_2017, darpacall_2023}, and the semiconductor industry \cite{moore2022ai, finkbeiner17}.}

\hpca{To further illustrate the efficacy of IMC, we contrast it with a hypothetical ASIC accelerator that operates at an identical speed and capacity. The aim is to match the integer multiplications of a Compute-Enabled Memory (CEM) as utilized in PPIMCE (See Section \ref{sec:IMC_Core}). In one operation (assuming a polynomial size N=8192), we would require 2$\times$8192 multipliers, which amounts to a total area of 343.6 $mm^2$ using data from \cite{vaidyanathan2015exploiting}. Conversely, a CEM only necessitates 4096 arrays, thereby only consuming an area of 138.5 $mm^2$. This indicates that IMC designs are around 2.5 times more area-efficient when achieving the same performance. Such efficiency underscores the effectiveness of IMC in managing the data movement overheads inherent in both HE and GC computations.}

\subsection{Challenges of Combining HE and GC}
\hpca{ PPIMCE aims to incorporate both HE and GC in a single IMC accelerator with the goal of supporting both HE and GC for the efficient execution of PPML tasks. However, unifying these two approaches in PPIMCE presents significant challenges due to the divergent computational and scheduling requirements.}

\hpca{\textbf{HE and GC computations fundamentally differ:}  HE computation is inherently multi-layered, e.g., neural network operations that devolve into HE arithmetic, polynomial arithmetic, and finally, coefficient-wise integer arithmetic\cite{GAZELLE}. In contrast, GC computation is based on two primary gates---AND and XOR---operating on GC labels~\cite{HalfGate,Freexor}. Especially, AND in GC entails AES encryption and other miscellaneous logical operations. Thus, HE and GC have distinct computational kernels, with HE leaning more towards integer arithmetic, and GC leaning on logical operations and AES encryption. PPIMCE reconciles these differences by employing IMC-cores for basic logical and arithmetic operations in/near memory for HE and GC (See Section \ref{sec:GC_basic},\ref{sec:HE_basic}).}

\hpca{\textbf{Scheduling Difficulties:} The scheduling requirements for HE and GC also diverge due to their distinct fundamental operations. In HE, coefficient-wise integer operations can typically be parallelized using Single Instruction, Multiple Data (SIMD) scheduling as these operations are mostly independent~\cite{HE_F1} However, in GC, the Boolean circuit (graph) demonstrates more significant data dependencies, which can vary across different tasks\cite{HAAC}. This variability makes it challenging to implement a universal scheduling mechanism as in HE. PPIMCE addresses this challenge by implementing a versatile scheduler that can effectively parallelize computation in HE while accurately tracking and managing data dependencies in GC (See Section \ref{sec:GC_mapping},\ref{sec:HE_mapping}).}

\section{Background}
\label{sec:background}
This section provides a brief introduction to HE and GC. For a complete description, see\cite{Yao,Freexor,HalfGate,Full_RNS_CKKS}.

\subsection{Homomorphic Encryption}
\subsubsection{HE basics}


Homomorphic encryption (HE) enables computation on encrypted data, and Fully Homomorphic Encryption (FHE) can theoretically evaluate any function. FHE schemes typically rely on the Ring Learning With Errors (RLWE) problem, using tuples of polynomials in the ring $R_q = \frac{\mathbb{Z}_q[X]}{X^N+1}$ for a power of two $N$. The B/FV and BGV FHE schemes work on finite-field plaintexts and are adaptable to machine learning applications \cite{B/FV,BGV}. The CKKS scheme \cite{CKKS}, which carries out approximate fixed-point arithmetic, is preferred for machine learning due to its native support for approximate arithmetic. Our work employs the CKKS scheme for its suitability to machine learning applications' approximate arithmetic \cite{kim2022approximate, lee2022privacy,lee2022low,BTS,chou2020privacy}. However, our architecture can support other FHE schemes due to its emphasis on improving the fundamental integer/polynomial operations used in FHE algorithms.

\subsubsection{Operations, Noise and Bootstrapping}

FHE schemes add noise to fresh ciphertexts in encryption, which accumulates as computations are performed \cite{B/FV,BGV,CKKS,TFHE}. Eventually, the noise becomes large enough that correct decryption is no longer possible. \textit{Bootstrapping} is a highly complex and expensive operation that reduces noise to tolerable levels without secret keys. In this work, we interpose GC between linear layers of neural networks; this has the additional effect of removing noise from ciphertexts\cite{GAZELLE,HE_cheetah}, obviating any need for us to perform bootstrapping. 

HE additions/multiplications are performed with sequences of polynomial additions, subtractions, multiplications, and scaling operations. HE rotations are performed by applying a polynomial automorphism to polynomials of the ciphertext and computing a dot product. All outcomes are closed in a polynomial ring, i.e., integer/polynomial modular reductions are performed after all the operations to keep the coefficients/degrees within a finite bound. For more details, please refer to the original schemes \cite{CKKS,Full_RNS_CKKS}.

\subsubsection{HE Optimizations}

Number-Theoretic Transform (NTT) and Residue Number System (RNS) serve as prominent algorithmic optimizations in Fully Homomorphic Encryption (FHE). NTT, by enabling polynomial multiplication in the evaluation domain to correspond to coefficient-wise multiplication in the original domain, lowers computational complexity and lets multiplication be executed in $O(N\log{N})$ time due to the log-linear time complexity of the NTT and its inverse \cite{NTT}. By keeping all polynomials in the evaluation domain in PPIMCE, expensive NTTs are reduced \cite{Full_RNS_CKKS,CryptoPIM}

Conversely, RNS optimization facilitates handling smaller coefficients in FHE polynomial calculations, enabling, for example, a polynomial with 512-bit coefficients to be represented as 16 polynomials with 32-bit coefficients. This method simplifies large computations and enables parallelization of all polynomial operations in hardware designs with ample computing resources \cite{RNS_original,RNS_justify}. The multi-core design of PPIMCE is well-positioned to take advantage of RNS for efficient computations.

\subsection{Garbled Circuits}
\label{sec:GC_background}
\subsubsection{GC basics}
\label{sec:GC_basic_background}

Garbled Circuits, introduced in 1986 \cite{Yao}, is a secure two-party computation scheme involving two key roles: the \textbf{Garbler} and the \textbf{Evaluator}. During the garbling phase, the Garbler encrypts Boolean circuits and prepares encrypted truth tables for all gates, which are then sent to the Evaluator \cite{GC_Prove}. The Evaluator uses these encrypted tables and inputs to process the GC during the evaluation phase.

To improve GC's performance, several optimizations, including Point-and-Permute \cite{Point}, Row Reduction \cite{Rowreduction}, FreeXOR~\cite{Freexor}, and Half-Gate \cite{HalfGate}, have been proposed. These reduce computation complexity and the size of garbled tables. The PPIMCE system leverages FreeXOR and Half-Gate as the basic operations for GC.

\subsubsection{FreeXOR and Half-Gate} FreeXOR allows secure execution of XOR gates without garbled tables \cite{Freexor}. Half-Gate optimizes the ciphertext size of the AND gate and, combined with FreeXOR, enables the construction of circuits with efficient garbled XOR and AND gates \cite{halfgate_better, HalfGate}.

The garbled tables generation in GC is handled through AES, which involves four primary operations: AddRoundKey, SubBytes, MixColumns, and ShiftRows \cite{AES}. These techniques collectively contribute to the effective operation of the GC in the PPIMCE system.
A more general introduction to GC can be found~\cite{GCgentleIntro}. 

\section{PPIMCE Architecture}
\label{sec:architecture}
\begin{figure}[t]
  \centering
  \resizebox{\columnwidth}{!}{\includegraphics{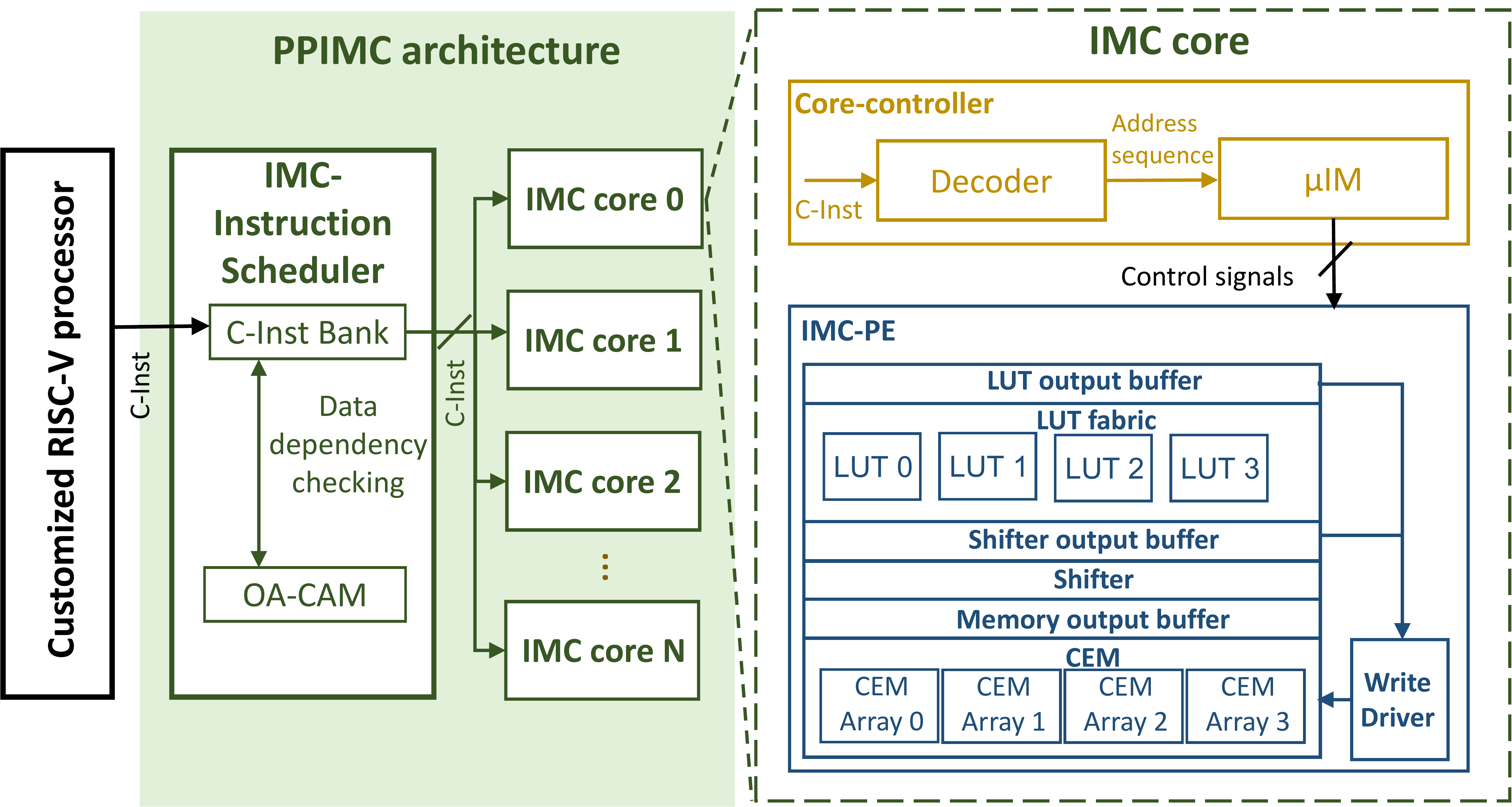}}
  \vspace*{-5mm}
  \caption{A high-level view of the PPIMCE accelerator and details of the IMC core. }
  \label{fig:overall_architecture}
\vspace*{-5mm}
\end{figure}

In this section, we introduce the architecture of PPIMCE. Figure \ref{fig:overall_architecture} shows the overall architecture of PPIMCE. PPIMCE consists of an IMC-Instruction Scheduler (IMC-IS) and multiple IMC cores. PPIMCE serves as a co-processor for a RISC-V processor extended with customized instructions (C-Inst) for HE and GC operations.  To execute an HE/GC operation, a C-Inst is issued to the IMC-IS. The IMC-IS dispatches the instruction to each core controller in each IMC core. The core controller decodes the C-Inst into control signals for computing units in the in-memory processing element (IMC-PE). The detailed architecture of each block is described below. Section \ref{sec:GC_HE_mapping} will discuss how PPIMCE performs HE and GC tasks.

\subsection{IMC-Instruction Scheduler}
\label{sec:scheduler}

 The IMC-IS dispatches  RISC-V instructions to IMC cores. The C-Inst Bank temporarily stores the C-Insts sent from the RISC-V, and the Output Address Content Addressable Memory (OA-CAM) checks the data dependency. CAM supports efficient parallel search \cite{CAM}. The OA-CAM and C-Inst Bank are set to 16KB, which is sufficient to handle PPIMCE data dependencies.

 IMC-IS employs the OA-CAM to ascertain data dependencies amongst instructions. An output address is deemed 'unavailable' if its instruction is being executed or waiting in the C-Inst Bank. Incoming instructions relying on these addresses must pause until prior instructions conclude. These unavailable addresses are held in the OA-CAM and are removed once the instruction is completed. Data dependencies are determined through an $O(1)$ time search in the OA-CAM using an instruction's input address \cite{CAM}. Both RISC-V instructions and those within the C-Inst Bank undergo this dependency check each cycle.

 PPIMCE's operational modes vary for HE and GC tasks. GC tasks require IMC cores to optimize parallelism. IMC cores can be grouped into a GC computing unit to parallelize GC gates and maximize hardware utilization. Instructions are dispatched to these units by the IMC-IS, and potential stalling scenarios are mitigated by storing RISC-V instructions in the C-Inst Bank until issues are resolved.

Conversely, HE tasks represented by a C-Inst can perform $N$ integer operations on each polynomial coefficient simultaneously across all IMC cores. As polynomial arithmetic in HE lacks data dependencies, instructions are dispatched to IMC cores without OA-CAM and Bank checks. Further details about IMC-IS functionality for GC and HE tasks are provided in Sections \ref{sec:GC_mapping} and \ref{sec:HE_mapping}, respectively.

\subsection{IMC Core}
\label{sec:IMC_Core}
There are multiple IMC cores in PPIMCE, and each IMC core contains several computing units in its IMC-PE as well as a core controller. We describe each component in the IMC core in detail and then illustrate how HE and GC's basic operations are mapped into the IMC core.

\subsubsection{IMC-PE}
The core component of the IMC-PE is the CEM, an innovative design adapted from IMCRYPTO \cite{imcrypto} that allows arithmetic and logic operations to be performed inside the SRAM array. However, the CEM is less efficient when handling permutation tasks, due to constant memory read/write operations, as well as LUT-based operations, as pre-storing LUT tables can compromise memory capacity. To overcome these inefficiencies, the IMC-PE is supplemented with a Shifter and a LUT fabric. These enhancements are tailored to better support the fundamental functions in both HE and GC (see Section \ref{sec:GC_basic} and \ref{sec:HE_basic}), optimizing the performance of the PPIMCE's IMC core.

The \textbf{LUT fabric} employs small memory elements (i.e., 6T-SRAM arrays and RA/CAM arrays of size 256$\times$8 \cite{imcrypto}) with customized peripherals, such as XOR networks (i.e., XOR trees). The memory elements of an LUT fabric and the XOR trees implement table-based multiplication over $GF(2^8)$, which is used in AES. The size of each memory element (i.e., 256$\times$8) is chosen so it is possible to store 256 pre-computed bytes (the size of an Sbox). In PPIMCE, besides AES, the regular 4-bit integer multiplication in HE is also implemented with pre-computed values stored in the LUT fabric. Each LUT fabric in an IMC-PE contains 4 RA/CAM arrays and 8 SRAM arrays, which enables a good trade-off between the multiplication speed for AES and HE implementations and the area overhead of PPIMCE.

The \textbf{shifter} of an IMC-PE performs byte permutations, rotations, and bit extensions. For instance, the ShiftRows (InvShiftRows) encryption steps (decryption) need byte permutations in AES. Note that only byte permutations were supported by the shifter in IMCRYPTO \cite{imcrypto}. Rotations and bit extensions were introduced in PPIMCE to support shift-add and integer reductions in HE and Half-Gates in GC (See Section \ref{sec:GC_HE_mapping} for more details).

Finally, the \textbf{CEM} comprises multiple arrays (called CEM arrays). With the aid of customized sense amplifiers, each CEM array can execute AND, OR, XOR, NOT, and ADD. Operations between two aligned memory words via the simultaneous activation of two wordlines. The size of a CEM array varies from tens of KB up to a few MB. A large CEM array can be useful when a CPU frequently reads cached data and sends it to external parties using communication protocols. On the other hand, large memories have longer access times and consume more power. To allow for a compromise between memory size, access times, and energy consumption, the CEM of a single IMC-PE is a 4 KB memory that consists of a tiled SRAM structure (with 4 tiles) that allows for the implementation of a high-throughput pipeline structure inside the IMC-PE. The CEM in PPIMCE is equipped with carry-lookahead adders, which can considerably improve addition time for long words (beneficial for HE).

\subsubsection{Core controller}
\label{sec:core controller}
To execute basic HE and GC operations more efficiently, we encode each HE and GC instruction with a sequence of micro-instructions and add a core controller to guide the execution of these micro-instructions in each IMC-PE. The micro-instruction execution is fully pipelined using core controllers. The micro-instructions are stored inside the micro-instruction memory ($\mu$IM). The size of $\mu$IM is set to 16 KB, which is sufficient to store all the micro-instructions needed for HE and GC.

Each micro-instruction is a 128-bit value that contains the control signals for each computing unit in IMC-PE. A 1-bit enable/disable and 1-bit memory mode switch signal are allocated for the LUT fabric. A 6-bit control signal (containing 1-bit enable/disable and 5 bits of function code) is included to specify the different shift operations (e.g., shift left, shift right, bit extension, etc.). Each CEM array has 1-bit enable/disable and 3-bit function codes for different in-memory computing operations and 26 bits for the corresponding memory addresses.

Each core controller also contains a decoder, which decodes a C-Inst into a $\mu$IM's address sequence. The $\mu$IM reads one micro-instruction out in each cycle until it reaches the end of the address sequence. The control signals in a micro-instruction are sent to the respective components in parallel.

\subsection{Customized RISC-V Processor}
\label{sec:RISCV}

In the PPIMCE architecture, a customized 32-bit RISC-V microprocessor is employed to efficiently manage HE and GC tasks. This involves enhancing the RISC-V ISA with ten new RV32I R-type instructions for HE and GC operations, while simultaneously updating micro-instructions and the LUT fabric accordingly. There are eight HE GC function instructions that execute fundamental HE and GC operations, including Half-Gate and FreeXOR for GC, polynomial manipulations (addition, subtraction, permutation, multiplication, reduction), NTT, and INTT for HE. Additionally, PPIMCE gains the capability to update micro-instructions in each core controller. A memory writes instruction that efficiently writes a micro-instruction to all core controllers simultaneously. By employing immediate values to define the micro-instruction and its address, it enables easy support for various HE and GC operations by adding new sequences. Furthermore, PPIMCE efficiently utilizes another RISC-V instruction to concurrently update content in all LUTs of each IMC core. 
\section{GC and HE mapping}
\label{sec:GC_HE_mapping}
This section details the execution of HE and GC's fundamental operations within the IMC core, and how PPIMCE manages HE and GC tasks.

\subsection{GC basic operations in IMC core}
\label{sec:GC_basic}
IMC cores perform Half-Gate and FreeXOR computations for GC tasks. The core controller performs static scheduling for dispatching the control signals of Half-Gate and FreeXOR into each component of the corresponding IMC core. Below, we describe how Half-Gate and FreeXOR are computed in the IMC core.

\textbf{Half-Gate:} The Half-Gate contains the AES-128 basic functions (AddRoundKey, SubBytes, MixColumns, and ShiftRows) \cite{AES} and other operations such as logical XOR, AND, and LSB extension of a label. The LUT fabric performs the SubBytes and MixColumns of AES in Half-Gate. The Shifter performs the ShiftRows and LSB extension. AddRoundKey and logic AND use in-memory XOR and AND in CEM arrays.

\textbf{FreeXOR:} FreeXOR can be performed via in-memory XOR in CEM arrays.

\subsection{HE basic operations in IMC core}
\label{sec:HE_basic}
The IMC core performs integer operations (integer reduction, integer addition, and integer multiplication) as the HE basic operations in polynomial computation for HE tasks. The core controller decodes integer operations into control signals and performs static scheduling to dispatch the control signals to each component. Below we describe how each integer operation in HE is computed in PPIMCE's IMC core.

\textbf{Integer reduction modulo $q$:} \rebuttal{In modern HE schemes, elements operate in the domain $R_q = \frac{\mathbb{Z}_q[X]}{X^N+1}$, where integer reduction modulo $q$ is applied to all coefficients as part of basic arithmetic operations. 
PPIMCE utilizes Barrett reduction \cite{Barrett} for general modular reduction. However, prior work has shown that choosing moduli of special form can bring performance improvements \cite{CIM_HE_SAC,wang2003moduli}.
Barrett reduction works with any modulus of any size, but it introduces considerable computational overhead due to the two integer multiplications it performs. 
PPIMCE can utilize a set of three specific moduli $q_i$ (with $q_0 = 2^{k}-1$, $q_1 = 2^{k}$, and $q_2 = 2^{k}+1$) as the ciphertext modulus $q$ for low-depth tasks like PPML inference, allowing optimizations for better performance. 
Conversely, Barrett reduction is employed for cases in our benchmarks with larger ciphertext moduli.}

\rebuttal{For optimized modular reduction after multiplication, we take as input an integer $X$ and produce $X_{q_i} = X \pmod{q_i}$ for $X \in [0, q_i^{2})$. }
We employ a similar algorithm described in \cite{CIM_HE_SAC} to avoid multiplication during the reduction process.
Initially, we calculate $X_{q_1}^\prime = X \land (2^{k}-1)$, $Y = X << k$, and $X^\prime = X_{q_i}^\prime + Y$. 
If $q = 2^{k}$, we directly use $X_{q_1}^\prime$ as the output. 
If $q_i = 2^{k} + 1$, we first check if $X_{q_1}^\prime \geq Y$ by performing $A = X_{q_1}^\prime - Y$ and extending the most significant bits of the temporary value $A$ to $A^\prime$ in the Shifter of the IMC core. 
This step checks the signed bits of $A$.
If $X_{q_1}^\prime < Y$, $A^\prime$ will have all 32 bits set to 1. 
Otherwise, $A^\prime$ will be 0. 
Next, we apply conditional logic using $A^\prime$ to choose the output between $X_{q_1}^\prime - Y$ and $X_{q_1}^\prime + (q_i-Y)$, where $X_{q_i} = (((X_{q_1}^\prime + (q_i-Y)) \oplus (X_{q_1}^\prime - Y)) \land A^\prime)\oplus (X_{q_1}^\prime - Y)$. 
If $q_i = 2^{k} - 1$, we perform a similar process to select the output between $X^\prime - q_i$ and $X^\prime$ based on the condition $X^\prime \geq q_i$.
When compared to executing Barrett reduction in the IMC core, this process can provide a 15\% performance improvement.

\textbf{Integer addition and subtraction:} Integer addition can be done using in-memory addition in CEM arrays. The CEM arrays do not support subtraction, but we can use NOT and addition operations to perform subtraction. We first perform a NOT operation on the subtrahend and store the result. Then we perform an ADD between the subtrahend and minuend and set the carry-in of the addition to 1.

\textbf{Integer multiplication:} Integer multiplication is the fundamental computing element of polynomial multiplication. Implementing integer multiplication in PPIMCE using the naive shift-add method requires $O(n^2)$ times (where $n$ represents the number of bits) shift-and-add operations in CEM arrays. A naive approach to this would lead to impractically high computation costs. To avoid this, we apply two optimizations in integer multiplications: \textbf{(i)} We use LUT fabrics to perform fast 4-bit integer multiplication; \textbf{(ii)} We employ the Karatsuba multiplication algorithm~\cite{karatsuba}.

We utilize the LUT fabric for 4-bit integer multiplication in a single clock cycle and employ the Karatsuba multiplication algorithm\cite{karatsuba} to recursively break down multiplications of two integers into multiplications of integers with half the number of bits. With a complexity of only $O(n^{1.6})$, the Karatsuba algorithm outperforms the naive approach. In PPIMCE, the base case for Karatsuba multiplication is 4-bit integer multiplication. The algorithm's addition operations are carried out using in-memory addition in CEM arrays of IMC cores, while left shift operations are executed in the IMC cores' Shifter.

\subsection{GC tasks in PPIMCE}

\label{sec:GC_mapping}
\begin{figure}[t]


  \centering
  \includegraphics[scale=0.178]{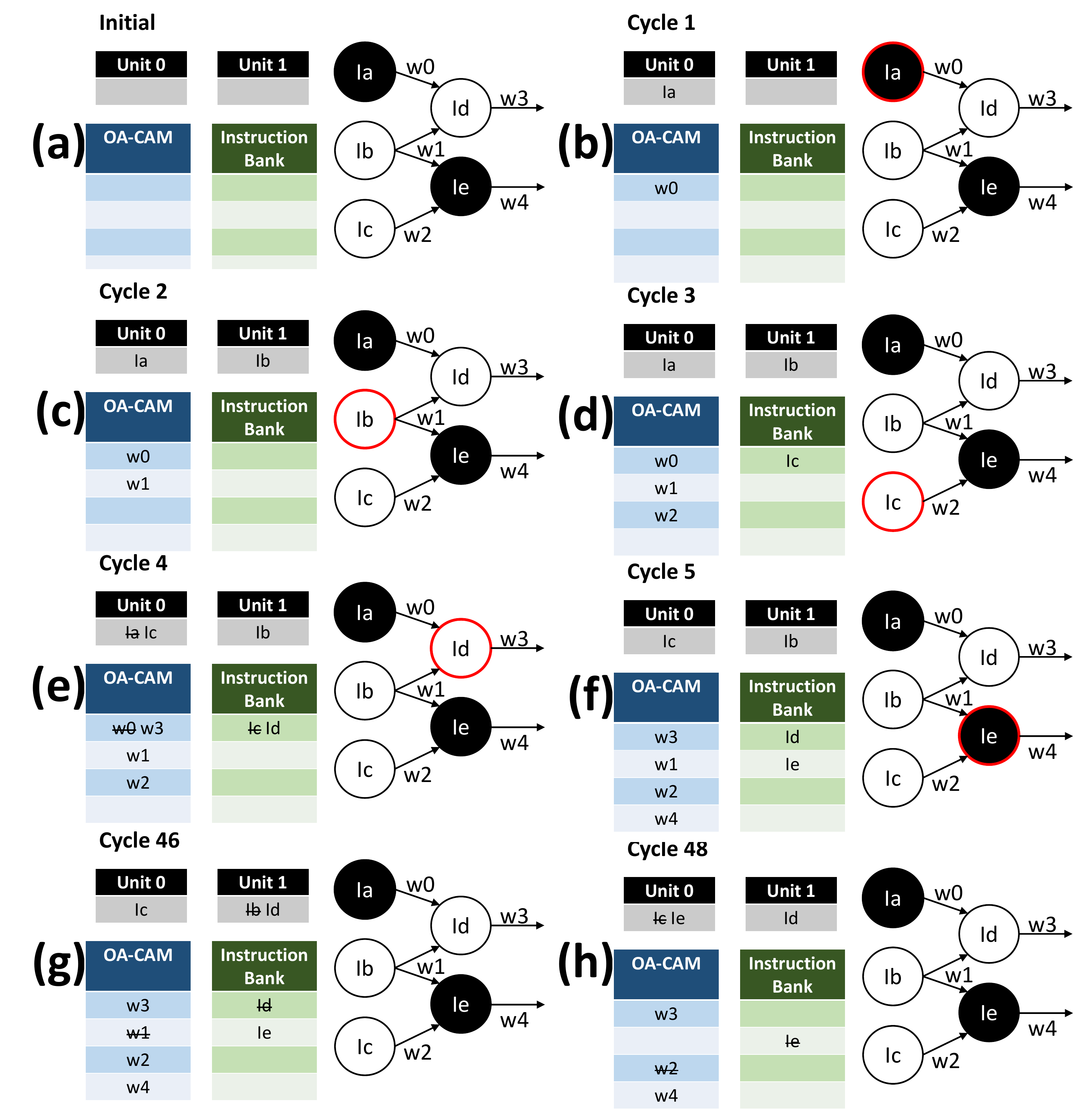}
  
\vspace*{-1.em}
  \caption{
  An example of GC instructions being executed on PPIMCE with only two GC computing units.
  The circle labeled with Ia -- Ie represents the C-Inst of Half-Gate (white cycle) and FreeXOR (black cycle), and the black arrows represent data dependency. $w0$ -- $w4$ are the gates' output. The instruction outlined in red represents the instruction fetched during the current cycle. The initial state is shown in (a). The states for cycles 1-5 are shown in (b)--(f). In cycle 4, Ia is completed. The states for cycles 46 and 48 are shown in (g) and (h), respectively, where Ib is completed in cycle 46, and Ic is completed in cycle 48.
  }

  \label{fig:GC_Mapping_example}
  \vspace*{-5mm}

\end{figure}

PPIMCE takes on the task of accelerating GC computation by first compiling GC tasks into customized Half-Gate and FreeXOR. We pre-generate and store necessary labels for table generation in the system's main memory. During the pre-processing phase, these labels are moved from the main memory to the CEM arrays to generate the garbled tables. PPIMCE also organizes multiple IMC cores into a GC Computing Unit, with each core operating in parallel, executing the same GC gates on different data. For instance, when executing a Half-Gate instruction on data stored at addresses 0 and 1, every IMC core performs the Half-Gate operation using data in their local address 0 and address 1. This coordination allows for efficient and parallel computation across all the cores in the GC computing unit.

In Figure \ref{fig:GC_Mapping_example}, we demonstrate how GC operations are dispatched to two GC computing units using OA-CAM and C-Inst Bank. In this example, we assume there are only two GC computing units. Each FreeXOR takes 3 cycles, and Half-Gate takes 45 cycles. The black and white circles represent the FreeXOR and Half-Gate gates, respectively.

In cycle 1, Ia is executed in unit 0, and $w0$ is written into OA-CAM. In cycle 2, Ib is executed in unit 1. In cycle 3, Ic is placed into C-Inst Bank, and $w2$ is written into OA-CAM. In cycle 4, Ia completes, freeing $w0$ and allowing Id to be written into C-Inst Bank, with $w3$ written into OA-CAM. Ic is issued into unit 0. In cycle 5, Ie is written into C-Inst Bank. In cycle 46, Ib completes, freeing $w1$ and enabling Id to be issued into unit 1. In cycle 48, Ic completes, releasing unit 0 and $w2$, allowing Ie to be issued into unit 0. At this point, all instructions have been issued to a unit.

\subsection{HE tasks in PPIMCE}
\label{sec:HE_mapping}




HE tasks consist of HE arithmetic including HE rotation, HE multiplication, and HE rotation. These operations are further broken down into polynomial arithmetic, which essentially comprises coefficient-wise integer computations. PPIMCE leverages its IMC cores, as detailed in Section \ref{sec:HE_basic}, to efficiently perform these integer computations in parallel.

In PPIMCE, we leverage its multiple IMC cores to store each coefficient of a polynomial at the same address within different cores. For example, the first coefficient is in address 1 of IMC core 1, the second coefficient is in address 1 of IMC core 2, and so forth. This setup enables a single address pointer to represent all coefficients in a polynomial. As a result, we can perform coefficient-wise arithmetic for polynomial multiplication, addition, and subtraction, all in parallel with a single command. When it comes to polynomial automorphism, PPIMCE handles read and write operations across the IMC cores to rearrange the coefficients in a polynomial, effectively accommodating the requirements of HE rotation operations.

 We propose a unique scheduling scheme for NTT and INTT operations in PPIMCE that requires only $N/2$ IMC cores to perform the butterfly computation \cite{NTT} on a polynomial of degree $N$. For instance, a PPIMCE with four IMC cores executing NTT on a degree-4 polynomial stores coefficients and corresponding twiddle factors evenly across the cores. Initially, coefficients $c$ and $d$ are moved to cores 0 and 1 for computing $cTW$ and $dTW$. Then, addition and subtraction operations occur in cores 0 and 1, producing temporary results $a'$ and $b'$ in core 0 and $c'$ and $d'$ in core 1. Next, cores 0 and 1 compute $b'*TW$ and $d'*TW$, respectively. Finally, the last butterfly computation is performed, placing the resulting polynomial coefficients in all four cores. This process involves only two IMC cores, enabling parallel execution of two NTTs with four cores and thus allowing for parallel execution of two NTT or INTT transformations on a degree-$N$ polynomial using $N$ IMC cores.

\section{PPML inference}
\label{sec:secure_inference}
\begin{figure}
  \centering
  \resizebox{\columnwidth}{!}{\includegraphics{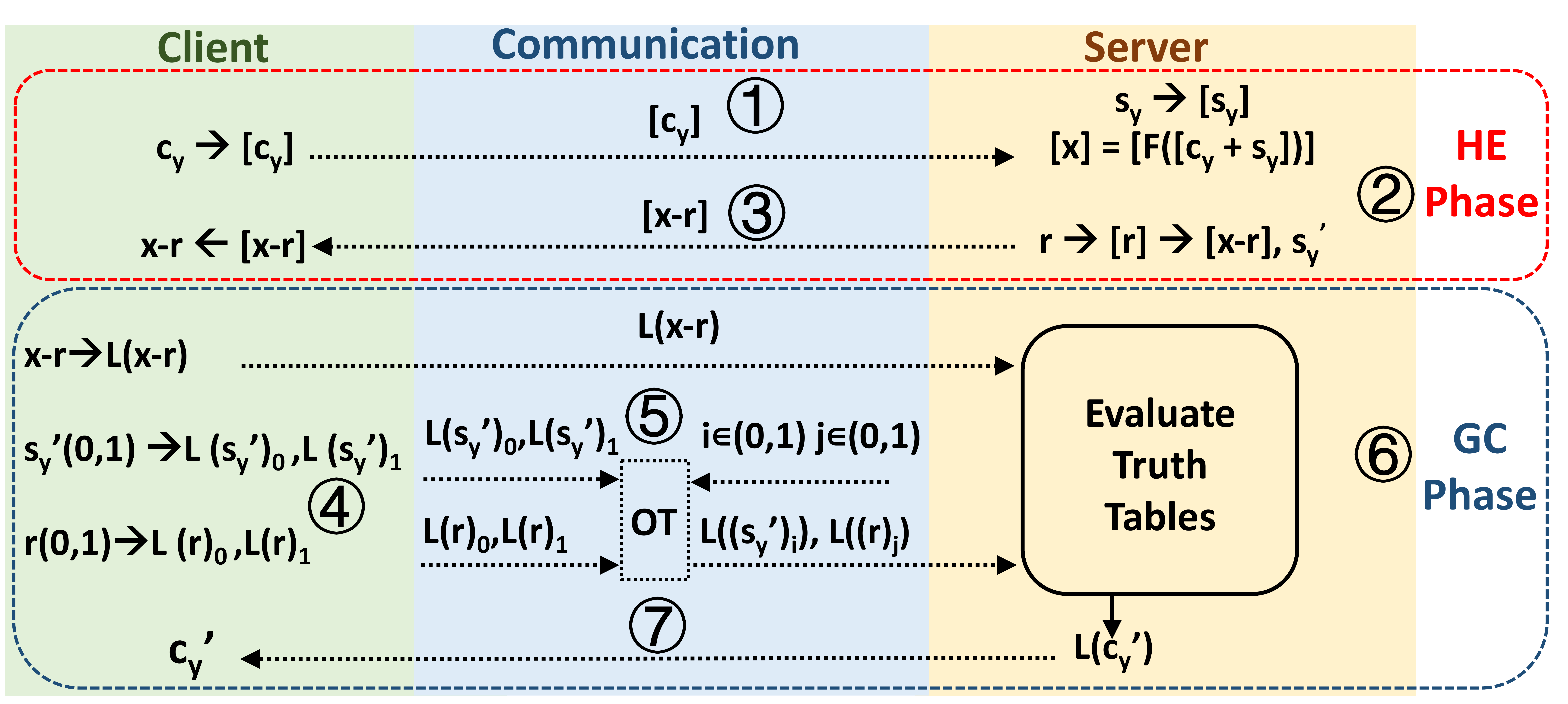}}
\vspace*{-5mm}
  \caption{The client-server architecture of PPIMCE for PPML inference. \textbf{[$\cdot$]} represents the ciphertext polynomial after homomorphic encryption. $L()$ represents the labels after GC label substitution. $F()$ represents the functions in the linear layer.}
  \label{fig:secure_ML}
\vspace*{-5mm}
\end{figure}

This section introduces a client-server architecture for PPML inference similar to \cite{GAZELLE}. The client holds the input data needed for inference, and the server holds the pre-trained network. This architecture aims to make the inference on the client's data without letting the server know the client's input data and without exposing critical data (e.g., client input data and server's network weights and bias) during communication. The PPML network contains linear layers (e.g., convolution and fully connected layers) and non-linear layers (e.g., ReLU and Maxpooling). The linear layers are computed using HE and the non-linear layers are computed using GC. PPIMCE can support both HE and GC, so there is no data transfer between the linear and non-linear layers of inference.

Figure \ref{fig:secure_ML} shows the client-server architecture of PPML. PPIMCE can operate either as a client or as a server. Below we detail the steps of the PPML protocol.  \circled{1} The client and the server first possess additive secret shares $c_{y}$ (on the client side) and $s_{y}$ (on the server side) of the linear layer input $y$, where $y=c_{y}+s_{y}$ (At linear layer 0, we set $c_{y} = y$ and $s_{y} = 0$). The client and the server encrypt $c_{y}$ and $s_{y}$ to a polynomial $[c_{y}]$ and $[s_{y}]$. The client sends $[c_{y}]$ to the server. \circled{2} The server executes $[x]=[F([c_{y}+s_{y}])]$ homomorphically (where $F()$ is the function in this linear layer). The server also subtracts a random value $r$ on $[x]$ to get $[x-r]$ homomorphically. 
This is to transform his ciphertext to additive secret shares.
The server also prepares a random number $s_{y}^\prime$ for the next phase.
\circled{3} The server sends $[x-r]$ to the client. The client uses HE decryption to get the value $x-r$. 
\circled{4} 
The client and server turn to the GC phase, where the client is the Garbler, and the server is the Evaluator. 
The value $x-r$, $r$, and $s_{y}^\prime$ 
are the inputs of the GC phase.
The client first picks label $L(x-r)$ corresponding to her own input $x-r$, and substitutes $L(r)$, $L(s_{y}^\prime)$ for all possible $r$ and $s_{y}^\prime$.
\circled{5} 
The client sends her label $L(x-r)$ directly, and lets the server picks his labels $L(r)$ and $L(s_{y}^\prime)$ via OT according to his own inputs.
\circled {6} 
The server evaluates the GC truth tables for $ReLU((x-r)+r)-s_{y}^\prime$.
The truth tables are independent of the inputs, so they can be stored on the server in the pre-processing phase~\cite{GAZELLE}.
The server uses the labels $L(x-r)$, $L(r)$ and $L(s_{y}^\prime)$ to evaluate the truth tables.
\circled{7} 
The evaluation result will be shared to the client to decode $c_{y}^\prime$ where $c_{y}^\prime = ReLU(x)-s_{y}^\prime$. 
The $c_{y}^\prime$ and the random value $s_{y}^\prime$ from the GC phase will transfer to the HE phase as the inputs $c_{y}$ and $s_{y}$ for the next HE phase.
Steps 1-7 are repeated for all the linear and non-linear layers until reaching the end of the network for the prediction result.
PPIMCE can transfer from HE to GC lightly on the client or server, as it supports both protocols in one implementation.

\section{PPIMCE Evaluation Setup}
\label{sec:implementation}

To validate the correct functionality of PPIMCE and evaluate its performance, including latency, power, and area, a comprehensive evaluation framework is indispensable. Toward this end, we develop a PPIMCE compiler, a PPIMCE cycle-accurate simulator, as well as a set of hardware simulators. Below, we describe how PPC tasks are executed in PPIMCE. 

\subsection{PPIMCE Evaluation Infrastructures}

  

 We leverage several existing tools at different abstraction levels to estimate the latency, energy, and area of PPIMCE for each basic GC and HE operation. Specifically, we have implemented C-Inst in the RISC-V processor in Verilog at the RTL level and evaluated it through detailed RTL simulations to ensure the correctness of the C-Inst fetching. The decoder in each core controller and the Shifter in each IMC-PE are also validated through RTL simulations. The LUT fabric and CEM arrays of each IMC-PE are validated at the circuit level with SPICE simulations. Finally, the DESTINY simulator \cite{destiny}, an open-source memory simulator, is used to estimate latency, area, and power for the C-Inst Bank in the IMC-IS and the $\mu$IM in each core controller. The latency, area, and power of the OA-CAM in the IMC-IS are measured using EVA-CAM \cite{eva_cam}, an evaluation tool for CAM. The area and power dissipation of PPIMCE includes the area and power of all the IMC-PEs, all core controllers, and the IMC-instruction schedulers. An IMC-PE's area and power dissipation consist of the area and power of the CEM arrays, the shifter, and the LUT fabric.

 We developed a PPIMCE compiler for compiling a PPC task described in C++ into a C-Inst list. The PPIMCE compiler includes the PPIMCE encoder and PPIMCE code generator. The PPIMCE encoder converts C++ code into HE and GC operations. For example, each linear layer of a DNN is converted into HE multiplications, additions, and rotations. Furthermore, the compiler converts each activation layer into GC ReLU operations. The PPIMCE code generator then generates the C-Inst list of polynomial arithmetic instructions for HE computation and Half-Gate and FreeXOR instructions for GC computation.

To estimate the delay and energy consumption of PPC tasks like PPML inference executed by PPIMCE, we developed a cycle-accurate simulator in Python 3.7 to evaluate PPIMCE's performance and explore the design space (e.g., the number of IMC cores). The simulator simulates the operations running in each IMC core cycle by cycle. Furthermore, the simulator meticulously tracks all data movement in the system, allowing us to account for all the data dependencies between the gates for GC functions and among the integer operations for HE functions (see Section \ref{sec:GC_HE_mapping}).

 \subsection{PPIMCE Parameter Setting}
 In the PPIMCE architecture, several parameters can significantly impact performance. We describe the trade-offs among the selection of values for these parameters.

Operations in HE, like integer multiplication, can be highly parallelized. For example, PPIMCE can parallelize all integer operations in HE functions using $N$ IMC cores if $N$ equals the polynomial degree. We choose PPIMCE with $8192$ IMC cores in our evaluation. $8192$ IMC cores allow PPIMCE to fully parallelize all operations in HE when $N=8192$, providing sufficient security levels and multiplicative depth for PPML inference.
 
The parallelism of GC functions is affected by the number of GC computing units in PPIMCE. We have studied all possible numbers of GC computing units to examine their impact on latency. The optimal number of units depends on the GC functions. Based on our study, we choose PPIMCE with 16 GC Computing Units, the Pareto optimal in terms of the number of units and latency for GC functions. Given the choice of $8192$, each GC Computing Unit contains $8192/16 = 512$ cores, allowing PPIMCE to run $512$ GC tasks in parallel.

 To study the impact of technology scaling on power and area, and to make a fair comparison with Cheetah\cite{HE_cheetah}, which uses 5nm nodes, we consider a 5nm technology node with foundry-reported scaling factors. Specifically, we use $0.079\times$ power and $0.059\times$ area to scale from 45nm to 7nm, based on \cite{Scaling45nm}. The power and area scaling factors are $0.70\times$ and $0.54\times$ from 7nm to 5nm, based on \cite{5nm}. Power and area scaling factors (45nm to 5nm) are $0.0553\times$ and $0.0318\times$, respectively.

Finally, we envision that PPIMCE will be placed on the same chip as the CPU, mimicking a last-level cache (LLC) to facilitate data exchange with the CPU. PPIMCE with 8192 IMC cores contains 32MB of on-chip memory, which may not be enough to hold all the data needed for a large-scale PPC task like PPML inference. PPIMCE needs to move the data from the main memory for computation. We assume 512 GB/s bandwidth between PPIMCE and the main memory (similar to HBM2 PHY bandwidth). PPIMCE executes HE functions in a computation-bound manner, allowing us to pipeline memory transfer and HE computation. On the other hand, the GC phase is memory-bound, but we can hide the memory transfer time in GC computation by pre-loading the data to PPIMCE's CEM arrays, such as the labels for the next computation during the current GC computation.

\section{Evaluation Results}
\label{sec:Experimental and Evaluation}

  
  

We first evaluate PPIMCE on GC and HE benchmarks and compares the results with CPU and GPU implementations. Then, we consider PPML inference and compare our design to existing PPC accelerators.  We use a computer with an Intel(R) Xeon(R) CPU E5-2680 v3 $@$ 2.50GHz for CPU evaluation and an NVIDIA RTX6000 for the GPU implementation. As existing GPU implementations for GC do not use exactly the same optimization as PPIMCE (Half-Gate, FreeXOR), we only compare PPIMCE with CPU implementation for GC evaluation. All components in PPIMCE are implemented in the 45nm CMOS predictive technology model (PTM) \cite{cao2011predictive}. We choose the operating frequency of 1 GHz for PPIMCE based on the longest basic operation in the IMC-PE.

\subsection{HE and GC benchmarks}

\begin{figure}[t]
  \centering
  \resizebox{\columnwidth}{!}{\includegraphics{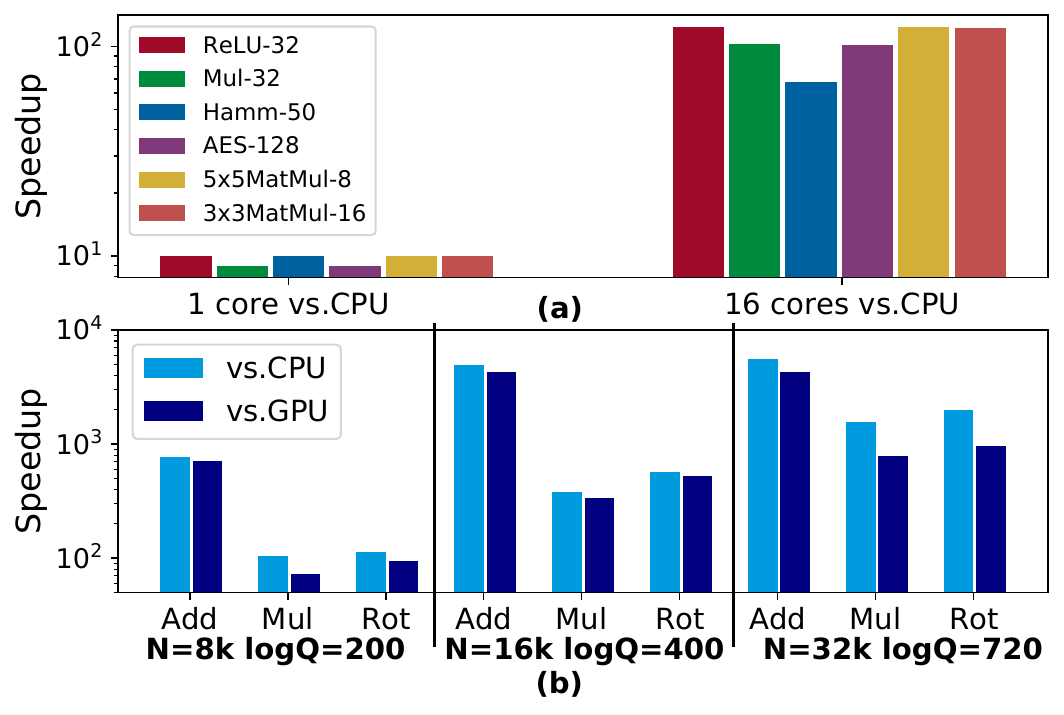}}
\vspace*{-8mm}
  \caption{
  (a) Speedup of PPIMCE with different numbers of IMC cores on GC benchmarks compared with a CPU implementation. 
  (b) Comparison between PPIMCE with 8192 cores on full RNS CKKS benchmarks (homomorphic addition (Add), homomorphic multiplication (Mul), and homomorphic rotation (Rot)) with CPU and GPU implementations for different HE parameters.
  }
  \label{fig:HE_GC_CPU_GPU_comparison}
\vspace*{-5mm}
\end{figure}

\subsubsection{GC performance}
We first evaluate the GC benchmarks in PPIMCE using the benchmarks from VIP-Bench \cite{VIP} and prior works~\cite{FASE,Maxelerator,Overlay}: \textbf{(i) ReLU-32:}  Perform an activation function to calculate $max(0, input)$ with 32-bit input size.   \textbf{(ii) Mul-32:} Perform a 32-bit integer multiplication with 32-bit output.   \textbf{(iii) Hamm-50:} Calculate the hamming distance between two 50-bit binary
values with 8-bit output.  \textbf{(iv) AES-128:} Perform a 128-bit AES encryption where each party separately provides the key and plaintext. \textbf{(v) 5$\times$5MatMul-8:} Perform a $5\times5$ matrix multiplication where each element in the matrix is 8 bits. \textbf{(vi) 3$\times$3MatMul-16:} Perform a $3\times3$ matrix multiplication where each element in the matrix is 16 bits.

\micro{A key challenge when comparing with GC benchmarks is the substantial on-chip memory space required to store temporary values. We employ four 1KB CEM arrays in each IMC core for area and power evaluation, which suffices for PPML GC non-linear functions and HE computation. 
However, this may not be sufficient for some GC functions, such as AES-128 and 5$\times$5MatMul-8, which require large memory space for intermediate data. This evaluation compares the PPIMCE speedup of GC microbenchmarks with a CPU realization. We increase each CEM array size in the IMC core to 128KB solely for GC microbenchmark evaluation, large enough to account for all GC benchmarks. This adjustment ensures a fair throughput comparison with CPU performance in this specific evaluation.}

Figure~\ref{fig:HE_GC_CPU_GPU_comparison}(a) reports the speedup of PPIMCE versus a CPU. 
PPIMCE performance is evaluated by scaling IMC cores from 1 to 16 with a fixed 128KB memory size. We compare the speedup of Garbler for generating the truth tables for the GC functions (i) - (vi). CPU-based GC is implemented with the EMP framework~\cite{emp-tool} for comparison. On the Evaluator slide, PPIMCE has a similar speedup. PPIMCE can achieve an average speedup of $9.6\times$ with a single IMC core compared with the CPU. The 16 IMC cores in the PPIMCE represent the Pareto-optimal solution that strikes a balance between area and latency. Because of data dependency in the GC, 16 cores cannot pump up the speed ideally to 16$\times$ compared with using 1 core. PPIMCE with 16 IMC cores achieves an average of 107$\times$ (10$\times$ faster than single-core PPIMCE) speedup. 






\subsubsection{HE performance}


Next, we conduct an evaluation of PPIMCE with 8192 IMC cores against CPUs and GPUs for three fundamental ciphertext-ciphertext operations in HE: HE addition, HE multiplication, and HE rotation. We use three different sets of HE parameters and the full RNS CKKS scheme. Due to the testing of high logQ values, Barrett reduction is employed in these operations. The performance comparison involves CPU and GPU projections utilizing the HEAAN library\cite{Full_RNS_CKKS}, a C++ library that implements the CKKS scheme exclusively, and a GPU-accelerated version using CUDA.

Figure \ref{fig:HE_GC_CPU_GPU_comparison} (b) illustrates the resultant speedup, indicating that PPIMCE can achieve a substantial speedup in operations. Specifically, $1500\times$ to $5000\times$ for HE addition, $100\times$ to $1500\times$ for multiplication, and $110\times$ to $2000\times$ for rotation. Although a GPU can manage a $2\times$ improvement for large parameter values compared to a CPU, PPIMCE still offers a speedup of up to $4000\times$, $800\times$, and $960\times$ for the same operations, respectively. 


    

\subsection{PPML inference}
\label{sec:PPML_inference}

\subsubsection{PPIMCE vs. Combined HE \& GC Designs}

We compare scenarios with IMC (PPIMCE) and without IMC (CraterLake+HAAC) using LoLA-CIFAR~\cite{lola}, a 6-layer secure ML model for CIFAR-10\cite{cifar-10} to highlight the efficiency of a uniform IMC accelerator. LoLA-CIFAR exclusively employs HE, replacing non-linear layers with square activation to accommodate HE. In contrast, we use GC for the non-linear layers and HE for the linear ones. We assume an ideal case where the control system of CraterLake+HAAC incurs no overhead, and the data transfer cost between the two accelerators is zero, allowing us to concentrate on the intrinsic computational performance and establish the performance upper bound for such a combined system.

CraterLake's LoLA inference performance includes linear HE computations and HE-friendly polynomial approximation functions for non-linear operations. To fairly compare PPIMCE with CraterLake+HAAC, we first isolate the time CraterLake spends on the linear layers. The reported data in~\cite{craterlake} does not provide the latency for only the linear layers. We compute this latency by analyzing the percentage of time spent on each layer in LoLA-CIFAR to extract the linear layers' latency. As approximately 80\% of the time is spent on linear layers, we estimate CraterLake's linear layer latency by multiplying its total LoLA-CIFAR time by 80\%.

Table \ref{tab:Combined_comparison} presents the performance comparison of PPIMCE and CraterLake+HAAC for LoLA-CIFAR inference. To ensure a fair comparison, we aim to maintain iso total latency for PPIMCE and CraterLake+HAAC and then compare their power dissipation and total area. It is challenging to manipulate CraterLake's design to match the HE latency of PPIMCE due to its more complex structure; however, we can achieve a similar GC performance with PPIMCE and with HAAC. Since HAAC is a smaller and more flexible design, we can employ multiple HAAC units for parallel computing, effectively matching the total latency of PPIMCE. We use 20 parallel HAAC units to parallelize the computation of non-linear functions in the combined system.

Respectively fabricated in 5nm, 16nm, and 45nm CMOS nodes, CraterLake, HAAC, and PPIMCE are scaled to a 5nm node for a balanced comparison using methods from \cite{Scaling45nm,5nm}. The rescaled PPIMCE is found to occupy significantly less area (138.3 $mm^2$) and consume less power (9.4 W) than CraterLake+HAAC (190.7$mm^2$ and 128 W). PPIMCE, thus, despite mirroring the total latency, exhibits a marked advantage in terms of area and power efficiency. These benefits are linked to PPIMCE's IMC computing approach, reducing data movement and facilitating the simultaneous acceleration of HE and GC.


\subsubsection{PPIMCE vs. Alternatives in PPML Inference}

\begin{table}[]
\centering
\caption{Comparison between PPIMCE and CraterLake+HAAC for LoLA-CIFAR Inference}
\label{tab:Combined_comparison}
\begin{tabular}{@{}l|c|c|c}
\hline
\hline
         & \textbf{PPIMCE} & \textbf{CraterLake} & \textbf{HAAC} \\ \hline
HE latency(ms)         & 52.3 & 40.4  & \textit{-} \\ 
GC latency(ms)        & 1.42 & \textit{-} & 13.9  \\ 

Area$^*$ ($mm^2$)         & \textit{-} & 157  & 33.7  \\ 
Power$^*$ ($W$)        &\textit{-} & 114.2 & 13.8 \\ \hline
Total latency ($ms$)        & 53.8 & \multicolumn{2}{c}{54.3} \\ 
Total area$^*$ ($mm^2$)            & 138.3 & \multicolumn{2}{c}{190.7}  \\   
Total power$^*$ ($W$)         & 9.4 & \multicolumn{2}{c}{128}  \\ \hline\hline
\multicolumn{4}{@{}p{\linewidth}@{}}{\small\textit{*All area and power are scaled to 5nm.}} \\

\end{tabular}
\vspace*{-5mm}
\end{table}

\begin{table*}[!b]
\vspace*{-5mm}
\centering
\small
\caption{Execution time (ms), accuracy, area ($mm^2$) and average power consumption (W) of PPIMCE, Gazelle \cite{GAZELLE} (CPU implementation), Cheetah\cite{HE_cheetah}, F1\cite{HE_F1}, BTS \cite{BTS}, CraterLake \cite{craterlake}, ARK \cite{ark} a single inference of ImageNet on ResNet-50 and CIFAR-10 on ResNet-20}

\label{tab:PPML_compare}

\begin{threeparttable}

\begin{tabular}{c |c c c c | c c c c| c c}
     \hline
     \hline
     & \multicolumn{4}{c|}{CIFAR-10 on ResNet20} & \multicolumn{4}{c}{ImageNet on ResNet50} \vline& \multirow{2}{*}{\begin{tabular}[c]{@{}c@{}} \\Area$^*$ \\ ($mm^2$) \end{tabular}} & \multirow{2}{*}{\begin{tabular}[c]{@{}c@{}}\\ Power $^*$ \\ ($W$) \end{tabular}} \\
     
     &\begin{tabular}[c]{@{}c@{}}HE time \\ (ms) \end{tabular} & \begin{tabular}[c]{@{}c@{}}GC time \\ (ms) \end{tabular} & \begin{tabular}[c]{@{}c@{}}Total time \\ (ms) \end{tabular} & Accuracy & \begin{tabular}[c]{@{}c@{}}HE time \\ (ms) \end{tabular} & \begin{tabular}[c]{@{}c@{}}GC time \\ (ms) \end{tabular} & \begin{tabular}[c]{@{}c@{}}Total time \\ (ms) \end{tabular} & Accuracy & & \\
     \hline
     Gazelle (CPU)  & 1.4e+4 & 3008 & 1.7e+4 & 91.9\% & 7.3e+6 & 1.3e+5 & 7.4e+6 & 76.1\% & - & - \\
    
     Cheetah  & - & - & -  & 91.9\% & 198  & 1.3e+5  &1.3e+5 & 76.1\% & 587 & 30\\
     F1  &2693 & 0 & 2693 & 90.7\% & -&- & -& LOW & 116 & 74 \\
     BTS  & 1910 & 0 & 1910 & 90.7\%  & - & - & - & LOW & 201 & 88 \\
     CraterLake  &249.4 & 0 &  249.4& 90.7\%  & - & - & -& LOW & 157 & 114.2 \\

     ARK  &125 & 0 &  125& 90.7\%  & - & - & -& LOW & 225.9 & 105 \\
     \textbf{PPIMCE} & \textbf{19.1} & \textbf{1.5} & \textbf{20.6}  & \textbf{91.9\%}& \textbf{7347} & \textbf{66.5} & \textbf{7413} & \textbf{76.1\%}& \textbf{138.3} & \textbf{9.4} \\
     \hline
     \hline

     \multicolumn{11}{@{}p{\linewidth}@{}}{\small\textit{*All area and power are scaled to 5nm}} \\
\end{tabular}
\end{threeparttable}

\end{table*}

\micro{Next, we compare PPIMCE with the SOTA software implementation, Gazelle \cite{GAZELLE}, as well as the SOTA PPC hardware accelerators including Cheetah\cite{HE_cheetah}, F1\cite{HE_F1}, BTS\cite{BTS}, CraterLake\cite{craterlake} and ARK\cite{ark} for end-to-end PPML inference. We compare these implementations' latency, accuracy, area, and power. We scale all designs' area and power to 5nm technology nodes for a fair comparison. Notice that all the HE discussed in this comparison utilizes ciphertext-plaintext arithmetic.}

We specifically focus on the server-side execution time, which comprises HE operations for linear layers and GC operations for non-linear functions in PPML inference (see Section \ref{sec:secure_inference}) \cite{cryptonite}. The protocols used by PPIMCE, Gazelle, and Cheetah are similar, leading to their total execution times being composed of both HE and GC times. For this analysis, PPIMCE adopts the same HE and GC parameters as Gazelle and Cheetah. Cheetah only accelerates the HE operations for PPML and does not accelerate GC. For a fair comparison, we assume that Cheetah uses the same GC process as Gazelle on the system's CPU for computations within activation functions. F1, BTS, and CraterLake only implement support for HE computation for PPML (with non-linear functions replaced by polynomial approximation\cite{HE_accuracy_dropping}), hence their execution time is solely comprised of HE computations.

This study assumes GC tables are transmitted during pre-processing (see Section \ref{sec:secure_inference}). Additionally, we assume that Cheetah, Gazelle, and PPIMCE have the same transmission requirements for PPML inference, as depicted in Fig. \ref{fig:secure_ML}. The transmission requirements for F1, BTS, CraterLake, and ARK are outlined in \cite{HE_resnet_20}. Our goal is to create a fair comparison between different PPML accelerators in terms of communication cost by making these assumptions.

We evaluated two PPML inference tasks: CIFAR-10\cite{cifar-10} on ResNet20 and ImageNet\cite{imagenet} for ResNet50\cite{Resnet}. Gazelle's execution time is measured by running its source code\cite{GazelleSource} on these two tasks. The performance data for other accelerators are obtained from their respective publications. Cheetah only reports the execution time for ImageNet on ResNet50. F1, BTS, CraterLake, and ARK only report execution time for CIFAR-10 on ResNet20.

 The accuracy of F1, BTS, CraterLake, and ARK performing CIFAR-10 inference on ResNet20 was reported in \cite{HE_resnet_20}. As these accelerators perform PPML exclusively with HE operations, inference accuracy drops — i.e., owing to the need for polynomial approximation, which accumulates the error during polynomial approximation\cite{HE_accuracy_dropping}. As such, while reasonable accuracy may be obtainable for networks for datasets such as CIFAR-10, accuracy is expected to plummet as more sophisticated networks/datasets are employed. PPIMCE, Cheetah, and Gazelle use GC for non-linear functions, so they do not have any accuracy drop. The accuracy of PPIMCE, Cheetah, and Gazelle for CIFAR-10 on ResNet20 is from\cite{HE_resnet_20}, and ImageNet on ResNet50 is from \cite{resnet50-acc}. 

 Table \ref{tab:PPML_compare} summarizes the performance results for PPIMCE and other accelerators. Compared with Gazelle, PPIMCE is not constrained by data transfer costs and can achieve high parallelism. PPIMCE can be up to 1000$\times$ faster than Gazelle. There are no performance improvements when using PPIMCE for HE versus Cheetah. However, despite fast computation for HE, Cheetah's total execution time is still impacted by GC computation overhead. PPIMCE obtains a 17$\times$ speedup compared to Cheetah.

 Compared with accelerators that use HE-only protocols, PPIMCE also gains significant speedup. The main reason is that over 95\% of the execution time in F1, BTS, CraterLake, and ARK on PPML inference is spent on bootstrapping (See Section \ref{sec:HE+GC_VS_HEonly}). With the support of GC for non-linear functions, PPIMCE does not incur the same high overheads for bootstrapping in PPML inference. PPIMCE can achieve 130$\times$, 90$\times$, and 12$\times$, 6.5$\times$ speedups versus F1, BTS, CraterLake, and ARK, respectively.

\rebuttal{PPIMCE surpasses other PP accelerators in terms of area consumption and power dissipation, chiefly due to its unique protocol and efficient In-SRAM IMC design. First, unlike existing HE-only accelerators like F1, BTS, CraterLake, and ARK which rely on computationally heavy bootstrapping\cite{HEProfiler}, PPIMCE adopts a Gazelle-like protocol that resets noise at every layer, avoiding bootstrapping. Second, based on previous research, in-SRAM computing design can offer approximately 2.5$\times$ energy and area savings compared to non-in-SRAM computing (See Section \ref{sec:Advantages_of_IMC}). The reason for the area-saving advantage is that computations occur at the bitline level, using customized sense amplifiers that only need a few extra transistors compared to conventional sense amplifiers. The energy-saving benefit is due to the fact that in-SRAM computing requires fewer external data accesses compared to regular non-in-SRAM computing.  Notably, the non-SRAM IMC solution Cheetah employs a similar protocol (GC-HE) to PPIMCE; Cheetah reports a power consumption of 30$W$, which is over 3$\times$ higher than PPIMCE, and in line with experimental results from  \ref{sec:Advantages_of_IMC} }
 
\begin{figure}

  \centering

  \resizebox{\columnwidth}{!}{\includegraphics{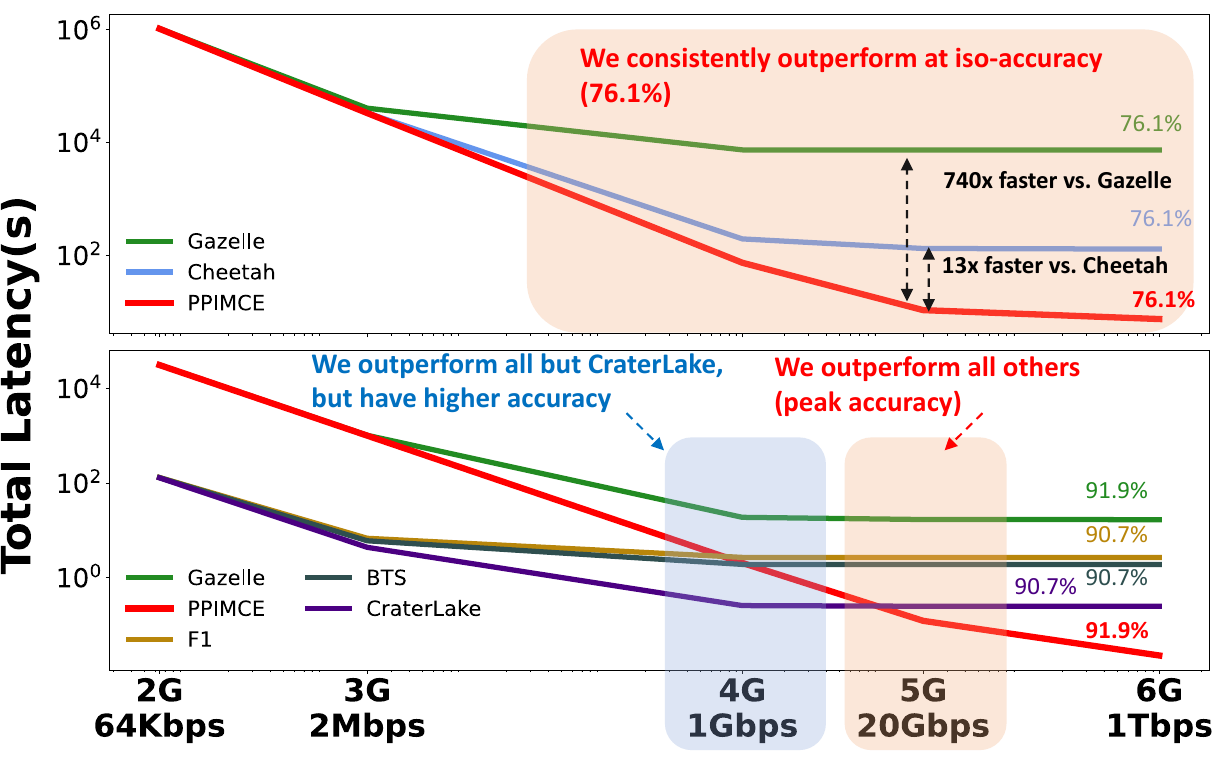}}

\vspace*{-1.em}
  \caption{Total latency of a single PPML inference vs. client-server communication bandwidth for PPIMCE and other implementations executing ResNet20 on CIFAR10 (lower) and ResNet50 on ImageNET (upper), followed by the note of inference accuracy.
  }
  \label{fig:HE_GC_comparison}
\vspace*{-5mm}
\end{figure}

\hpca{Finally, we analyze how the primary bottleneck of GC, the client-server communication, impacts the runtime of PPML inference in PPIMCE. By merging computation data from Table \ref{tab:PPML_compare} with communication latency, we illustrate the total PPML inference time for all designs in Figure \ref{fig:HE_GC_comparison}. We calculate communication latency utilizing the bandwidth of wireless protocols from 2G to 5G from ITU recommendation \cite{2G,3G,4G,5G}, and potential 6G bandwidth \cite{6G,6G_above}. The communication demand in PPIMCE is notably high; specifically, it requires 2GB for single inference on ResNet20, while the HE-only protocol only needs 8MB. With increasing bandwidth, communication time is reduced until computation time becomes the dominant factor, marking a saturation point. In low bandwidth scenarios, HE-only accelerators perform better in total latency, yet PPIMCE and other HE+GC accelerators achieve higher accuracy. In high bandwidth circumstances, PPIMCE surpasses all competitors in either latency or accuracy.}

\section{Related Work}
\label{sec:Related Work}

\textbf{HE acceleration:}
HE accelerators F1\cite{HE_F1}, CraterLake\cite{craterlake}, BTS\cite{BTS} and ARK \cite{ark} reduce HE computational overhead. F1\cite{HE_F1} performs well on various HE tasks but only supports low multiplicative depth. CraterLake \cite{craterlake},  BTS\cite{BTS}, and ARK \cite{ark} offer unbounded multiplicative depth but are limited to HE tasks, and unable to handle complex PPML tasks with high accuracy.

Existing IMC designs for HE reduces data-transfer overhead. CIM-HE\cite{CIM_HE,CIM_HE_SAC}, CryptoPIM\cite{CryptoPIM} and X-poly\cite{Xpoly} are accelerators performing HE arithmetic and logic operations in SRAM,  Resistive RAM, and crossbar, respectively, focusing solely on HE.

\textbf{GC acceleration:} Hardware accelerators for GC aim for high throughput with minimal area and power overhead. Recent FPGA implementations \cite{GC_overlay}\cite{Garbled_cloud} speed up Yao's GC, but lack advanced optimizations. Maxelerator \cite{Maxelerator} is an FPGA GC accelerator for matrix multiplication. FASE \cite{FASE} is the current SOTA FPGA GC accelerator with a deeply pipelined architecture and optimized scheduling.

\textbf{PPML acceleration:}
Several efforts have been made to use HE and GC to design specialized protocols for various applications. Software accelerators like Gazelle\cite{GAZELLE} and Delphi\cite{Delphi} use HE and GC to speed up PPML tasks. Gazelle uses HE for linear functions and GC for non-linear functions, while Delphi has a similar architecture but uses pre-processing to reduce communication costs. Cheetah \cite{HE_cheetah} is an ASIC-based hardware accelerator that adapts Gazelle's framework but only accelerates the HE part and requires additional support for the GC part, resulting in additional overheads.

\section{Conclusion}
\label{sec:Conclusion}

We propose PPIMCE, the first IMC accelerator for HE and GC that enables high throughput while reducing data transfer overheads. PPIMCE achieves significant speedup, up to 100 $\times$ compared to GC CPU implementations and up to 1500$\times$ and 800$\times$ speedup compared to CPU and GPU implementations when executing CKKS-based homomorphic multiplications. PPIMCE accelerates PPML inference using HE and GC without sacrificing accuracy and achieves up to 1000$\times$ speedup on single image inference compared to the SOTA CPU implementation Gazelle. Moreover, compared to the best-performing PPC accelerators, PPIMCE achieves speedups of up to 130$\times$ and exhibits superior area and power efficiency.



\bibliographystyle{IEEEtranS}
\bibliography{ref}

\end{document}